\documentclass{aa}  

\usepackage{graphicx}
\usepackage{txfonts}
\usepackage{aas_macros}
\usepackage{color}			
\usepackage{breakurl}		

\usepackage{xcolor} 
\usepackage{ulem} 

\begin{document}

\title{A survey on Hungaria asteroids involved in mean motion resonances with Mars}
	\author{
		E. Forgács-Dajka\inst{1,2,3} 
		\and 
		Zs. S\'andor\inst{1,2,4} 
		\and 
		J. Sztakovics\inst{1,5}
		}
	\institute{
		Department of Astronomy, Institute of Geography and Earth Sciences, E\"otv\"os Lor\'and University,\\
		H-1117 Budapest, P\'azm\'any P\'eter s\'et\'any 1/A, Hungary\\
		\email{e.forgacs-dajka@astro.elte.hu; zs.sandor@astro.elte.hu}
		\and
		Centre for Astrophysics and Space Science, E\"otv\"os Lor\'and University,\\
		H-1117 Budapest, P\'azm\'any P\'eter s\'et\'any 1/A, Hungary
		\and
		Wigner Research Centre for Physics, P.O. Box 49, Budapest H-1525, Hungary
		\and
		Konkoly Observatory, Research Centre for Astronomy and Earth Sciences, Budapest, Hungary
		\and
		Eszterházy Károly Catholic University, Faculty of Natural Sciences, Institute of Chemistry and Physics, Department of Physics, H-3300, Eger, Leányka utca 6-8.\\
		\email{sztakovics.jan@uni-eszterhazy.hu}
		}

\titlerunning{Dynamical survey on resonant Hungarias }

\date{Received ; accepted }

 \abstract
   {A region at the inner edge of the main asteroid belt is populated by the Hungaria asteroids. Among these objects, the Hungaria family is formed as the result of a catastrophic disruption of (434) Hungaria asteroid hundred million years ago. Due to the Yarkovsky effect, the fragments depending on their direction of rotation are slowly drifting inward or outward from the actual place of collision. Due to this slow drift these bodies could approach the locations of the various mean-motion resonances (MMRs) of outer type with Mars.}
   {We aim to study the actual dynamical structure of Hungaria asteroids that is primarily shaped by various MMRs of outer type with Mars. Moreover, we also seek connections between the orbital characteristics of Hungaria asteroids and their absolute magnitude.}
   {To map the resonant structure and dynamics of asteroids belonging to the Hungaria group, we use the method FAIR (as FAst Identification of mean motion Resonances), which can detect MMRs without the a priori knowledge of the critical argument. We also compile stability maps of the regions around the MMRs by using the maximal variations in the asteroids' eccentricities, semi-major axes, and inclinations. We numerically integrate the orbits of all asteroids belonging to the Hungaria group available in the JPL Horizon database together with the Solar System planets for one and ten million years.}
   {Having studied the resonant structure of the Hungaria group, we find that several asteroids are involved in various MMRs with Mars. We identify both short and long-term MMRs. Besides, we also find a relationship between the absolute magnitude of asteroids and the MMR in which they are involved.}
  {}


   \maketitle
%

\section{Introduction}\label{sec:intro}

Asteroids located in the main belt (orbiting between Mars and Jupiter) show highly complex dynamical behaviour. The most spectacular manifestation of this complex dynamics is the existence of the Kirkwood-gaps that are minima in the number distribution of asteroids with respect to their semi-major axes at the locations of the 3:1, 5:2, 7:3, and 2:1 mean motion resonances (MMRs) with Jupiter.   

In the innermost region of the main belt there is a group of asteroids that occupies a dynamically protected region that is practically neither affected by the MMRs with Jupiter, nor the most important secular resonances. This is the group of Hungaria asteroids (named after (434) Hungaria) whose members orbit within a narrow range of semi-major axis ($1.78-2.06$ au) being separated from the majority of the main asteroid belt by the 4:1 MMR of inner type with Jupiter. This region is dynamically bounded by the $\nu_6$, $\nu_{16}$ secular resonances on the $(a,i)$ plane \citep{2010Icar..207..769M}. Hungaria asteroids revolve in highly inclined orbits ($i=16^\circ-30^\circ$), while their eccentricities range between low to moderate values ($e < 0.18$). The higher limit in their eccentricity is close to that value being necessary for asteroids to cross the orbit of Mars \citep{1979aste.book..359G,2010Icar..207..769M}.

The present Hungaria asteroids could be the survivors of the much more extended asteroid belt, the hypothetical "E-belt", that has been destabilized by the late migration of giant planets, see \citet{2012Natur.485...78B}. Members of the Hungaria group are also unique in the sense that they are the closest asteroids to the Earth on dynamically stable orbits. Due to their peculiar position in the inner Solar System, and relative proximity to the orbit of Earth, the dynamics of Hungaria asteroids is of high interest. Investigations revealed that the members of Hungaria population are slowly escaping by crossing the orbit of Mars, therefore they are thought to be unstable over the age of the Solar System, \citep{2010Icar..207..769M,2010Icar..210..644M}. It has been found recently that the eccentricity of Mars, that is changing chaotically on a very long timescale, plays an important role in affecting the stability of Hungaria asteroids \citep{2018Icar..304....9C}. The higher eccentricity of Mars enhances the probability of close encounters of the asteroids, therefore the destabilization of Hungaria asteroids is rather not due to secular effects. This latter result may make a bit uncertain the conclusions of the previous studies on the determination of the half life time and decline rate of Hungaria asteroids. Moreover, there is another escape mechanism through chaotic diffusion of Hungaria asteroids. That is when due the Yarkovsky effect \citep{1999Sci...283.1507F} asteroids are drifted towards the locations of MMRs with Mars, where due to to the combined effects of the resonant and non-conservative perturbations, their eccentricities are gradually increased. \cite{2010Icar..210..644M} have found by numerical integration that the eccentricities of Hungaria asteroids involved in MMRs can increase until reaching the Mars orbit crossing limit. As a consequence, the chaotic variation of the eccentricity of Mars could also have a strong impact on the destabilization of the Hungaria asteroids that are involved in various MMRs with Mars. (Here we note that the Yarkovsky effect is a non-conservative force that is due to the solar irradiation and thermal emission of a rotating small asteroid. This effect manifests as a slow orbital drift that depends on the direction of the asteroid's rotation.)

According to dynamical studies on Hungaria asteroids, the boundaries of Hungaria asteroids can further be refined: they are clustered between the 4:3 and 3:2 MMRs of outer type with Mars, while the inclinations of these objects are bounded by $\nu_5$, $\nu_4$ linear, and $\nu_{56}$ nonlinear secular resonances, see for instance \cite{2018MNRAS.479.1694C}. A region of strong chaotic behaviour and instability develops in the $(a,e)$ parameter plane already at lower values of eccentricities that is the result of overlapping of various MMRs of outer type with Mars \cite{2018MNRAS.479.1694C}. Thus in that work a further evidence is given on the importance of dynamical effects of Mars through MMRs that seem to be essential when the stability boundaries of Hungaria asteroids are investigated. Regarding the investigations related to the MMRs between Hungaria asteroids and Mars it is noteworthy to mention the pioneering work of \citet{2008Icar..194..789C}, in which asteroids have been identified in the 3:2 MMR. In a preliminary work, \cite{2019LPI....50.2674S} investigated the MMRs between all known Hungaria asteroids and Mars with the method FAIR \citep{2018MNRAS.477.3383F}. Finally, we mention the work of \citet{2021Icar..36714564C}, in which the authors compiled a catalog of Hungaria asteroids being involved in the 3:2 MMR with Mars. It is also noteworthy to mention the work of \citet{1997Icar..128..230M} on the secular resonances in the inner Solar System those locations mark the boundaries for Hungaria asteroids.

Another important result on Hungaria asteroids is the identification of the Hungaria collisional family among the background population \citep{1994ASPC...63..140L}. This finding has further been confirmed by \citet{2009Icar..204..172W} and \citet{2010Icar..207..769M}, assuming that the Hungaria collisional family was formed by a catastrophic disruption of (434) Hungaria approximately 300 Myr or 500 Myr ago. A collisional family is usually spreading out, because beside the gravitational perturbations, fragments formed at the disruption event are also subject to the Yarkovsky effect. As the result of the orbital drift of asteroids occurred by the Yarkovsky effect, family members are gradually departing from the location of the disruption event. Smaller fragments that are more sensitive to the Yarkovsky effect, are drifted to larger distances from the location of the disruption than larger fragments. The zone that can host family members can thus be determined by plotting the V-shaped curved lines being characteristic for a collisional born family, as seen in the work of \citet{2009Icar..204..172W} for Hungaria asteroids. A reliable description on the formation of the above mentioned this V-shape distribution can be found in \cite{2019MNRAS.484.1815P}. Interestingly, the Yarkovsky effect may not work when an asteroid is captured into a MMR \citep{2015aste.book..509V}.

A more accurate distinction between family members and background asteroids can be done by calculating their proper elements. In the case of Hungaria asteroids, proper elements have been calculated first by \citet{1994ASPC...63..140L}, later on the "synthetic" proper elements in large number have been determined by \cite{2010Icar..207..769M}. More recently, \cite{2019MNRAS.484.3755V} determined the proper elements of more than 650000 asteroids also including the Hungaria asteroids. The proper elements in this work have been determined by an empirical method, in which the calculation of the secular perturbations is based on the distribution of the orbital elements. This method enables one to distinguish between the two most important effects, namely the classical perturbation and the Lidov-Kozai effect. 
According to the author, in the case of Hungaria asteroids neither the method based on the classic secular theory nor the empirical method give reliable results that is due to the MMRs that can be found in this region.

According to the investigations of \cite{2019Icar..322..227L}, one can also make a distinction between Hungaria family members and background asteroids. Based on the relative taxonomic distribution it has been found that 71\% of the background population belong to the S-type, while 24\% of them belong to the X-type. On the contrary, 88\% of the Hungaria family members are classified as X-type, while the remaining asteroids belong to the C and B-type. We should note, however, that in this case according to \cite{2019MNRAS.484.3755V} the taxonomical composition for the Hungaria family has not been changed. The C/X-type asteroids in the Hungaria family are of the E type that is the consequence of the very high albedos of the asteroids. Moreover, the C, X, and E types are spectrally very similar to each other therefore they might be confused.

Based on the above considerations, in our survey we do not distinguish between the Hungaria collisional family members and background asteroids, instead of that we globally study all asteroids that share the orbital elements that are characteristic to the Hungaria group. We also note that we investigate the dynamical behaviour of the Hungaria group members that is mainly governed by various MMRs of outer type with Mars, as it is reflected in the structure of our paper. 

In the second section we describe the asteroid data and their source. We use two data sets originating from two different epochs, which enables us to compare the effects of the observations to the results obtained in our dynamical survey. In this section we briefly describe the methods applied, namely the method FAIR \citep{2018MNRAS.477.3383F} that is our main tool of identifying the different MMRs, and also its automatisation making possible to detect MMRs on the large sample of asteroids included in our survey. In the third section we analyse our numerical results. We compare the asteroids exhibiting resonant behaviour in both data sets. 

Similarly to previous works \citep{1985Icar...63..272W,2021Icar..36714564C} we introduce the concepts of the short-term and long-term resonances based on the length of the time interval during an asteroid is involved in a certain MMR. Using the numerical results obtained, we compile dynamical maps of certain parts of the whole region filled by Hungaria group members to better understand the dynamical effects of the various MMRs. Beside the well known, classical method of the maximum variation of asteroids' eccentricities, we also use the maximum variation of other orbital elements (as the asteroids' semi-major axes and inclinations) that help obtain a comprehensive picture regarding the variety of MMRs in this region. Finally, we statistically investigate a possible relationship between the location of Hungaria asteroids involved in MMRs and their absolute magnitude using observational data. The most plausible reason of this size segregation effect could be the Yarkovsky effect, however other dynamical effects could also play an important role here. 

\begin{figure}
   \includegraphics[width=0.9\linewidth]{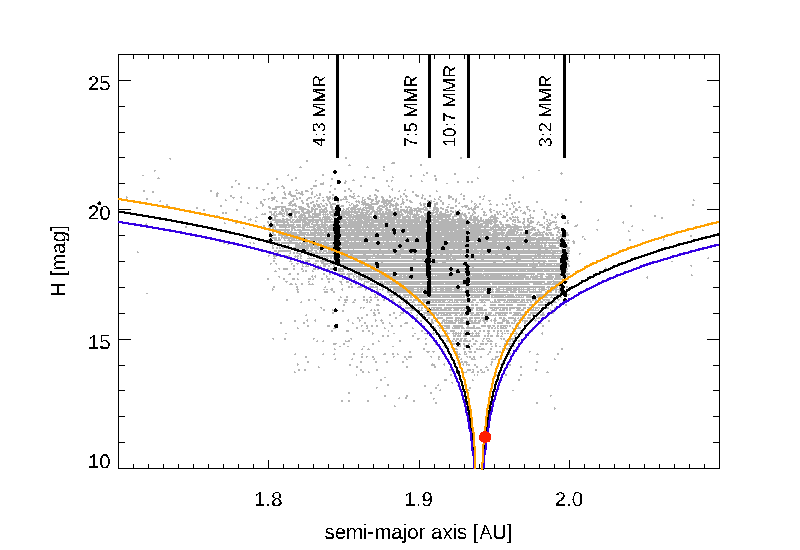}
   \caption{Hungaria-population asteroids projected onto the plane of absolute magnitude H versus semi-major axis of nearly 24000 objects. The lines correspond to the four most populated MMRs with Mars. Asteroid (434) Hungaria is marked by a red point. The curved lines show Hungaria-family zone, where we used a canonical form $0.2H = \log((a-a_c)/C)$, where $a_c = 1.94\mathrm{au}$ and $C=[3\times 10^{-5},2.5\times 10^{-5},2\times 10^{-5}]\mathrm{au}$, denotes by [blue,black,orange] and discussed by \cite{2006Icar..182..118V} and \cite{2009Icar..204..172W}.
   }
   \label{fig:sma-H}
\end{figure} 

\begin{figure*}
  \centering
  \includegraphics[width=0.9\linewidth]{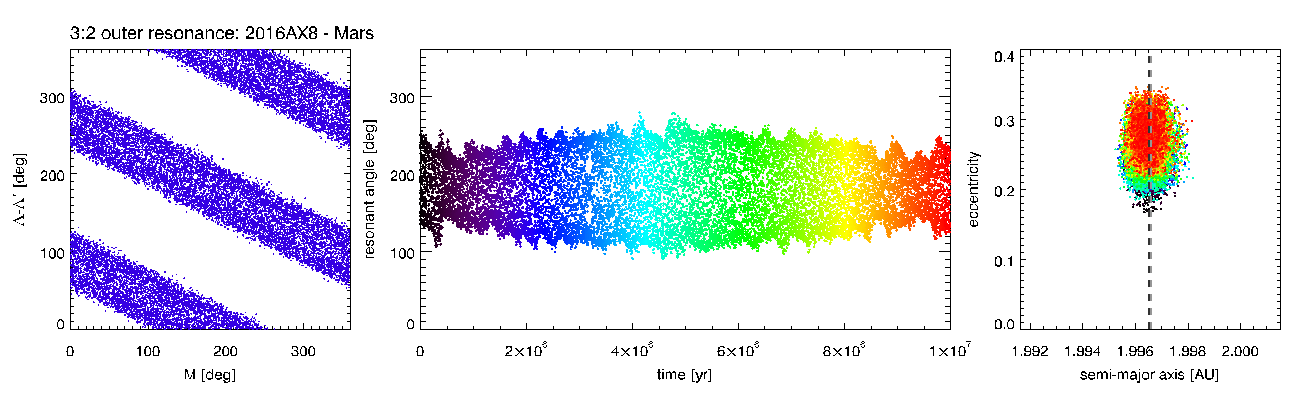}
  \caption{The dynamics of the asteroid $2016_{\mathrm{AX}8}$ in 3:2 MMR of outer type with Mars. In left panel method FAIR indicates the 3:2 MMR. In the middle panel one can see the variation of the critical argument showing libration for ten million years. In the right panel the excursion of the orbit on the $(a,e)$ is shown. The colour coding is explained in the text.}
\label{fig:res-2016AX8-mmr32}
\end{figure*}

\begin{figure*}
  \centering
  \includegraphics[width=0.9\linewidth]{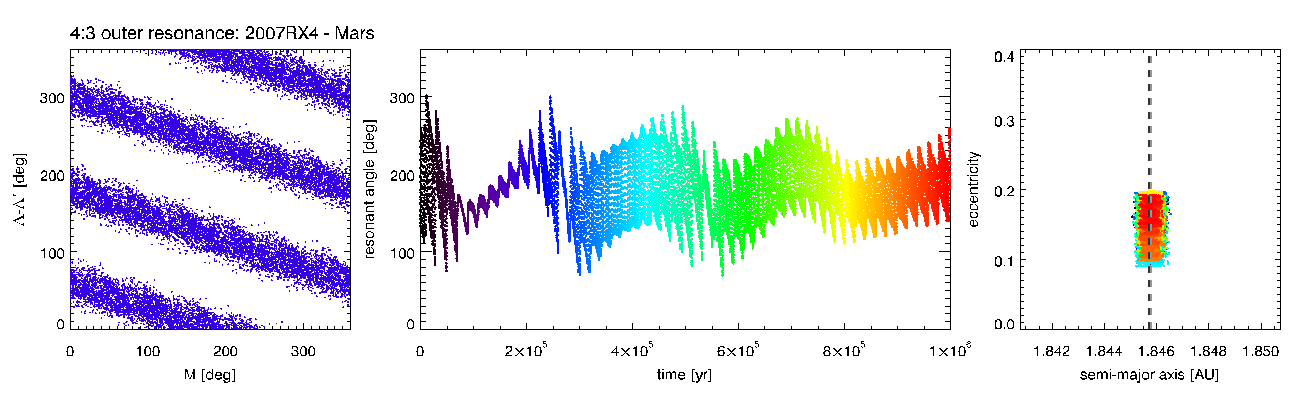}
  \caption{The same as in Fig.~\ref{fig:res-2016AX8-mmr32}, but the asteroid $2007_{\mathrm{RX}4}$ in 4:3 MMR and the integration time is one million years.}
\label{fig:res-2007RX4-mmr43-1E6}
\end{figure*} 

\begin{figure*}
  \centering
  \includegraphics[width=0.9\linewidth]{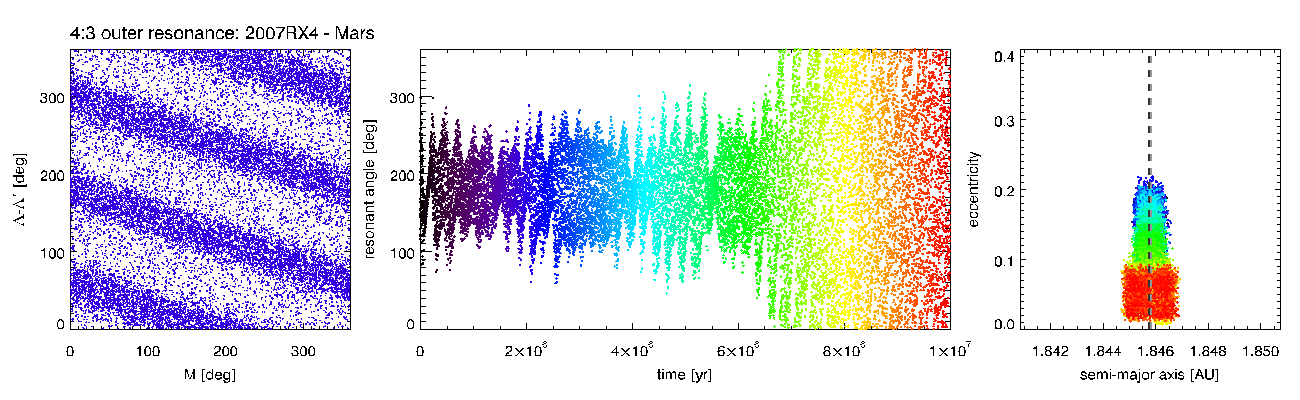}
  \caption{The same as in Fig.~\ref{fig:res-2007RX4-mmr43-1E6}, but the integration time is ten million years.}
\label{fig:res-2007RX4-mmr43-1E7}
\end{figure*} 

\begin{figure*}
  \centering
  \includegraphics[width=0.9\linewidth]{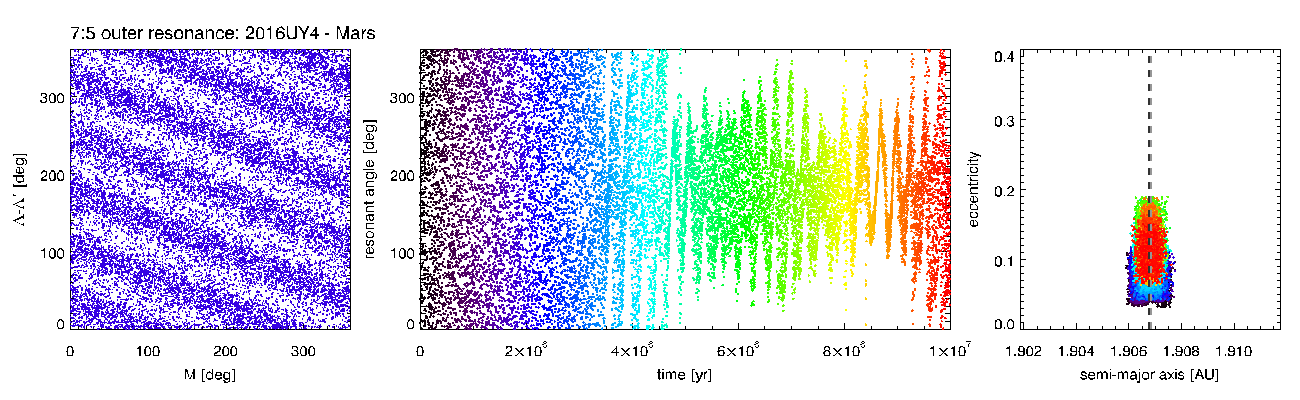}
  \caption{The same as in Fig.~\ref{fig:res-2016AX8-mmr32}, but the asteroid $2016_{\mathrm{UY}4}$ in 7:5 MMR.}
\label{fig:res-2016UY4-mmr75}
\end{figure*} 

\section{Data and methods}\label{sec:DataMethod}

Due to the recent sky surveys the number of objects with orbit type of Hungaria asteroids is rapidly growing, in 2019 the number of such asteroids was 19529, while in 2020 this number changed to 23860 being a significant increase of 22\%. In our work we consider those objects that are classified as Hungaria orbit type by IAU Minor Planet Center\footnote{https://www.minorplanetcenter.net/iau/mpc.html}, while the heliocentric coordinates and velocities of all identified bodies are taken from NASA JPL Horizon\footnote{https://ssd.jpl.nasa.gov/?horizons} database.  

As a first step of our data analysis, we display all members of the Hungaria group (found in the above database) in the $(a,H)$ plane, where $a$ is the semi-major axis and $H$ is the absolute magnitude of asteroids. As expected, most of the asteroids are located within the curved V-shaped lines on the $(a,H)$ plane (Fig.~\ref{fig:sma-H}), whose canonical form are given by \cite{2006Icar..182..118V}, indicating that they are in the Hungaria family zone. The parent asteroid of the family, (434) Hungaria is denoted by a red dot. We note that Fig.~\ref{fig:sma-H} is similar to the corresponding figure in \citep{2009Icar..204..172W}, however, we also display by thick black dots all asteroids involved in MMRs of outer type with Mars. (The automatised procedure of identification of asteroids being involved in MMRs is described later on in this section.) From Fig.~\ref{fig:sma-H} one can immediately conclude that the 4:3, 7:5, 10:7, and 3:2 MMRs, all of them of outer type with Mars, are the most populated ones. An interesting feature of Fig.~\ref{fig:sma-H} is the sharp limit of Hungaria asteroids at the 3:2 MMR with Mars, where a relatively large number of asteroids are affected by the resonance. At the 4:3 MMR, a considerable number of Hungaria asteroids are involved, too, however, the inner boundary of the Hungaria population seems to be located at the 9:7 MMR, where only a few asteroids are identified. While the sign of the Yarkovsky effect is clearly seen in the figure, another interesting feature is the asymmetry in the distribution of asteroids: a significant number of fainter objects are orbiting at lower semi-major axes, moreover, their orbital distance is larger from the orbit of (434) Hungaria, which is the progenitor of the collisional family, than the orbital distance of asteroids at the outer boundary of the whole Hungaria population. One possible explanation of the asymmetrical distribution that is compatible with the Yarkovsky effect, is given in Section 4 with the help of a detailed statistical analysis.

As already mentioned, stability studies have proposed so far natural boundaries of the stability region hosting Hungaria asteroids \citep[for example][]{2010Icar..207..769M,2018MNRAS.479.1694C}. Moreover, there are several MMRs with Mars that may govern the dynamics of Hungarias inside their stability region, \citep[see][for details]{2010Icar..210..644M}. In order to reveal the role of these MMRs, we numerically integrate the orbits of all members of the Hungaria group found in database, together with all major planets of the Solar System. The numerical integrations have been calculated by using an own-developed adaptive step size Runge-Kutta-Nystrom 6/7 N-body integrator \citep{1978CeMec..18..223D}. We note that our work is the first comprehensive dynamical survey of Hungaria asteroids so far aiming at the identification of each object being in MMR with Mars.

Having integrated the orbits numerically we apply the method FAIR (as FAst Identification of mean motion Resonances) to quickly detect all asteroids involved in MMRs with Mars. In what follows we shortly describe the method FAIR, but its detailed description and application can be found in \cite{2018MNRAS.477.3383F}. Let us denote by $n$ the mean motion, $\lambda$ the mean longitude, and $\varpi$ the longitude of perihelion of the asteroid, while the same primed quantities ($n^\prime$, $\lambda^\prime$, and $\varpi^\prime$) refer to the perturber (in our case to Mars). Considering an outer type MMR, the ratio of the mean motions satisfies
\begin{equation}
    \frac{n^{\prime}}{n} \approx \frac{p+q}{p},
\end{equation}
with $p$ and $q$ positive integers. According to FAIR, when displaying $\lambda - \lambda^{\prime}$ versus $M$ there appear stripes that intersect the horizontal axis in $p$ nodes while the vertical axis in $p+q$ nodes. The critical argument for this MMR
\begin{equation}
    \theta^{(p+q):p} = (p+q)\lambda - p\lambda^{\prime} - q\varpi
    \label{eq:crit_arg}
\end{equation}
librates around the centre of the resonance. Thus the method FAIR can be ideally applied to a large sample of bodies that members might be involved in various MMRs with an external perturber without the a priori knowledge of the librating critical argument. The above described application of FAIR can be seen in Figure \ref{fig:res-2016AX8-mmr32}, where the left panel displays the stripes on the $\lambda - \lambda^{\prime}$ versus $M$ plane, while in the middle panel the critical argument librates as a function of time. The colours indicate the time evolution in order to reveal the dynamics on the semi-major axis versus eccentricity $(a,e)$ plane shown in the right panel. In this panel we denote the resonant semi-major axis with dashed vertical line, and it is clearly visible that the $a,e$ values cover a quite large region. 

To detect whether a Hungaria asteroid is in a MMR with Mars, we automatise the use of the method FAIR. We recall that the number of nodes along the x and y axes should be detected that are the crossings of the stripes with the horizontal and vertical axes. To do so, we create a two dimensional distribution of all points on the $\lambda - \lambda^{\prime}$ versus $M$ plane by covering it by a bi-dimensional grid, and counting the number of points falling in each cell. The binsize we use is 5 degrees in both directions. We note that although this choice seems to be arbitrary, according to the naked eye checks it gives reliable results in identifying MMRs. After calculating the average of the points on this plane, we have determined the variance in each cell. If the variance is larger than a predefined threshold then the examined body might be in a MMR. In the next step one should determine the crossing of the stripes with axes, as we already mentioned. We cut along both x and y directions two not too narrow rectangular regions (we used 20 degrees) and create two histograms in both directions. We apply a Fast Fourier Transform (FFT) on these histograms that gives the number of nodes, e.g. the values of $p$ and $q$ in \eqref{eq:crit_arg}. As a last step of this method we check whether the MMR detected is around the nominal position of the resonance. To test this method, we check the results by human eye. The above method finds somewhat less cases (approximately by 10 percent) than detected by naked eyes. This apparent error of this method is due to the fact that there are two temporal types of MMR, i.e. long-term and short-term resonances (see later for a more detailed explanation), and in the case of very short-term resonances the variance is very small. It is important to note that in the case of the short-term resonances the sampling frequency affects the detection of a MMR because rarely sampling makes difficult to identify the stripes on the $\lambda - \lambda^{\prime}$ versus $M$ plane.

\section{The resonant structure of the Hungaria asteroids}

In this part we summarize our findings obtained by a dynamical and statistical survey of nearly 24000 asteroids belonging to the Hungaria group. 

In the first phase of our research we compare the differences that occur by using the databases from years 2019 and 2020. As we mentioned before, the number of asteroids is increased, on the other hand the orbital elements of the asteroids might be more accurate as the result of new and continuous observations. We investigate how the population and lifetime of a MMR are modified by the possible changes of the orbital elements due to the two different sets of orbital data. In some cases the formerly librating critical argument does not librate according to the new database from year 2020, or the long-term libration becomes temporary or in other words short-term libration.  

In the second phase we investigate how the integration time affects the population of asteroids involved in various MMRs, and also the time length that an asteroid spends in a MMR. By using the orbital elements in the database from year 2020, we integrate numerically the orbits of asteroids together with the Solar System planets for 1 and 10 million years. Some of the asteroids that are involved in a MMR for 1 million years showing only libration may loose their resonant character in the 10 million years long integration. This is a clear signature of the chaotic behaviour that is the result of the resonant perturbations from Mars and non-resonant ones from Jupiter and other planets. To explore the dynamical character of asteroids, we also compiled various stability maps of some of the most populated MMRs.

\subsection{Comparison of dynamics of asteroids from databases of years 2019 and 2020}

\begin{table}[]
    \centering
    \caption{The abundance of the various MMRs taken from the 2019 and 2020 databases.}
    \begin{tabular}{c|c|c|c|c}
        MMR   & SMA  & \multicolumn{3}{c}{No. asteroids}  \\
              & [au] & 2019 & 2020 & Both \\
        \hline  
        &&&\\
         7:6  &   1.689 &   -    &  1   &   0 \\
        14:11 &   1.789 &   -    &  2   &   0 \\   
         9:7  &   1.802 &   8    &  4   &   3 \\   
        13:10 &   1.815 &   2    &  2   &   0 \\   
         4:3  &   1.846 &   84 (28\%)   &  110 (30\%) &   58 \\   
        15:11 &   1.874 &   4    &  9   &   2 \\   
        11:8  &   1.884 &   5    &  11  &   3 \\   
        18:13 &   1.893 &   -    &  2   &   0 \\   
         7:5  &   1.907 &   85 (28\%)  &  112 (30\%) &   61 \\   
        17:12 &   1.922 &   5    &  4   &   1 \\   
        10:7  &   1.933 &   26  (8\%)  &  34 (9\%)  &   17 \\   
        13:9  &   1.947 &   7    &  5   &   2 \\   
         3:2  &   1.997 &   75  (25\%)  &  67  (18\%)  &   48 
    \end{tabular}
    \tablefoot{In the first column the MMR, in the second one the exact value of resonant semi-major axis, the third and fourth columns the number of resonant asteroids (with librating critical argument) from the 2019 and 2020 databases, respectively. In the last column are the numbers of resonant asteroids in both databases. In the 2019 database the number of Hungaria asteroids is 19530, while in the 2020 database this number is 23861. We show in parentheses the percentage of asteroids captured in the corresponding MMR among the whole population of resonant asteroids.}
    \label{tab:compMMRtwodata}
\end{table}

In Table~\ref{tab:compMMRtwodata} we show how the abundance of resonant Hungaria asteroids changes by using the orbital elements in the two databases of years 2019 and 2020. We can state that in the database from year 2020, the increase in the number of resonant asteroids is approximately $20$\% that corresponds to the change in the whole population of Hungaria asteroids. In the database from year 2019 we find 301 resonant asteroids, while in the  database from year 2020 this number is increased resulting in 363 resonant asteroids. Based on the above fact, both databases can be considered for statistical purposes. On the other hand, by comparing the two databases we can discover interesting differences, too. 

We expect that since the number of known asteroids is increased by 20\%, the number of resonant asteroids also increases by the same amount, that is confirmed by the results obtained. On the other hand, if we carefully investigate the distribution of asteroids captured in MMRs, we can conclude that in the 4:3, 7:5, and 10:7 cases the number of asteroids is increased, while those captured in the 3:2 MMR is decreased. If we consider the ratios of librating objects in a certain MMR to the whole resonant population, the distribution of asteriods in certain MMRs is also changing, see Table~\ref{tab:compMMRtwodata}. The reason of this fact might be the consequence of observations. Additionally, in the last column of the table the numbers of those resonant asteroids are shown that librate in both databases. Curiously, in the $3:2$ MMR the number of librating objects is proportionally larger than in the other MMRs. The reason of this difference could be that the orbital elements are known more accurate in this case.

\begin{table}[]
    \centering
    \caption{The same as the Table~\ref{tab:compMMRtwodata}, but here the number of asteroids captured in MMR during different integration times is shown.}
    \begin{tabular}{c|c|c|c|c}
        MMR   & SMA  & \multicolumn{3}{c}{No. asteroids}  \\
              & [au] & one million yrs & ten million yrs & Both \\
        \hline  
        &&&\\
         7:6  &   1.689 &  1           &  1           &  1       \\
        14:11 &   1.789 &  2           &  1           &  -       \\   
         9:7  &   1.802 &  4           &  4           &  3       \\   
        13:10 &   1.815 &  2           &  1           &  1       \\   
         4:3  &   1.846 &  110 (30\%)  &  112 (34\%)  &  76      \\   
        15:11 &   1.874 &  9           &  1           &  -       \\   
        11:8  &   1.884 &  11          &  14          &  5       \\   
        18:13 &   1.893 &  2           &  -           &  -       \\   
         7:5  &   1.907 & 112 (30\%)   &  105 (32\%)  &  57      \\   
        17:12 &   1.922 &  4           &  -           &  -       \\   
        10:7  &   1.933 & 34  (9\%)    &  32  (10\%)  &  19      \\   
        13:9  &   1.947 &  5           &  2           &  1       \\   
         3:2  &   1.997 &  67 (18\%)   &  57  (17\%)  &  39      
    \end{tabular}
    \label{tab:compMMRtwoyears}
\end{table}

Similar to the previous cases, where we examined the differences due to the initial conditions, we also examined how the integration time affects the population of a given MMR and the distribution of resonant asteroids in different resonances. The results are shown in Table~\ref{tab:compMMRtwoyears}, where the number of asteroids caught in each resonance is given by the two integration times, and the last column shows the number of celestial bodies that were resonant in both runs. Here we note that in this case, we started from the 2020 data as an initial condition. 
Comparing the two integration times, we can see that the total number of resonant celestial bodies decreased (for one million years 363 asteroids, for ten million years 330 ones), and small fluctuations can also be observed in the distributions. We note here that each asteroid showing libration in some interval of the whole integration time is already considered a resonant celestial body. This results in an interesting phenomenon, for example, that in the case of 4:3 MMR, the number of resonant asteroids increased during the longer integration time, because those that did not show resonant behaviour on a short time scale could be captured on a longer time scale (after the first one million years). In almost all of the main MMRs (except the 3:2 one) the number of asteroids does not decrease significantly after the ten million years long integration. Interestingly, the number of asteroids involved in the 3:2 MMR decreased a lot. The role of this MMR as a natural outer boundary of the Hungaria group, and the chaotic diffusion that may be acting here is discussed further in Section~\ref{sec:conclusion}.

If we examine those asteroids that show libration on both integration time scales, we see that the number of these asteroids is less than the number obtained during one million years of integration. The reason for this is to be found in the sampling frequency and the length of the libration period. This is because those asteroids that were only librating for a short time interval during the one million years integration may not have shown a detectable variation in the $\lambda - \lambda^{\prime}$ versus $M$ plane at the sampling frequency of the ten million year integration (cf.~Section~\ref{sec:DataMethod}). 

\subsection{Long-term and short-term MMRs between Hungaria asteroids and Mars}

In the following, we would like to present the two types of resonant behaviour that we detect in the case of the Hungaria asteroids. The difference between these two cases is the length of the time interval on which the critical argument shows libration. Although the semi-major axis of Mars does not change significantly, perturbations mainly from the giant planets can cause that the capture of asteroids into MMRs becomes episodic, which is why we observe that in many cases the critical argument librates for a shorter time than the whole length of numerical integration. In some cases, the resonance turns out to be long-lasting, as the critical argument shows libration during the ten-million-year integration time. These cases are classified as long-term MMRs. We note, however, that even the long-term libration of the critical argument can be changed into circulation if we integrate the orbit for a longer time. As already mentioned, the FAIR method is suitable for the detection of both types of resonance (short-term and long-term), taking into account the sampling time. We note that episodic libration of the critical argument in the case of the 3:1 MMR has been first observed by \citet{1985Icar...63..272W}, and very recently in the case of Hungaria asteroids involved in the 3:2 MMR of outer type with Mars by \citet{2021Icar..36714564C}.

Since the above cited paper deals with Hungaria asteroids involved in the 3:2 MMR of outer type with Mars, we could compare our results with the results of \citet{2021Icar..36714564C}. The authors identify 22 asteroids whose critical arguments librate during the 2 million years long numerical integration, see Table 1 in \citet{2021Icar..36714564C}. A majority of these bodies can also be found in our database, all have librating critical argument, except one asteroid (referred as 458733 in \citet{2021Icar..36714564C} and $2011_{\mathrm{NN}2}$ in our database). In Table 3 of \citet{2021Icar..36714564C} there are asteroids referred as "quasi-resonant". Among these bodies we find some of them with librating critical argument for the whole length of numerical integration. We have to note, however, that there are a few sources of discrepancies between our work and the work of \citet{2021Icar..36714564C}: (i) we use different numerical integrator, (ii) there could be differences in the initial conditions used, (iii) moreover asteroids are selected to the databases used in both studies by using different criteria.

In Fig.~\ref{fig:res-2016AX8-mmr32} we show the case in which the asteroid 2016$_{\mathrm{AX}8}$ is involved into the 3:2 long-term MMR with Mars. On the left panel the stripes clearly indicate the character of the resonance, while in the middle panel the libration of the corresponding critical argument, that is calculated based on the left panel. Finally, on the right panel the corresponding values of the semi-major axis and eccentricity are shown around the nominal value of the resonant semi-major axis, $a_r$. In the middle and right panels a colour coding is applied: the black dots indicate the onset, while the red dots indicate epochs close to the end of the numerical integration, the intermediate colours are for intermediate epochs. With this coding, the scattering of the $a,e$ values around the nominal value of the MMR can be followed. In this particular case the corresponding values are changing in time, for example it is clearly seen that larger libration amplitude of the critical argument implies larger variations in $a$.

In Figs.~\ref{fig:res-2007RX4-mmr43-1E6} and \ref{fig:res-2007RX4-mmr43-1E7} the short-term variation of the critical argument of the asteroid 2007$_{\mathrm{RX}4}$ is shown, in the case of the 4:3 MMR with Mars for one million years and ten million years long integrations, respectively. During the one million years long integration the critical argument only librates showing slightly irregular behaviour, but without any sign that later on this libration stops. In the case of the ten million years numerical integration the libration of the critical argument changes to circulation around $7\times 10^6$ years. Interestingly, libration occurs for relatively larger eccentricity values, in this particular case between 0.1 and 0.2, while when the critical argument circulates the eccentricity is confined between 0.01 and 0.1. Moreover, the libration of the critical argument is accompanied with smaller variations in the asteroid's semi-major axis than its variation when the critical argument circulates.

An interesting example for short-term resonance can be seen in Fig.~\ref{fig:res-2016UY4-mmr75}, where the asteroid 2016$_{\mathrm{UY}4}$ is in 7:5 MMR with Mars. In the beginning of the numerical integration the critical argument circulates, around $5\times 10^6$ years it turns to libration, while close to the end of integration it circulates again. Similarly to the case of the asteroid 2007$_{\mathrm{RX}4}$ (see Fig.~\ref{fig:res-2007RX4-mmr43-1E7}), when the critical argument librates, the corresponding $a,e$ points (coloured by green and orange) are confined between larger eccentricity values, while the variations of the semi-major axis are closer to the exact value of the MMR. We emphasize that if the semi-major axis of an asteroid is near to a certain resonant semi-major axis would not mean automatically that the body is captured in that MMR. The asteroid is captured in a MMR only if the corresponding critical argument librates.

\begin{figure*}
  \centering
  \includegraphics[width=0.45\linewidth]{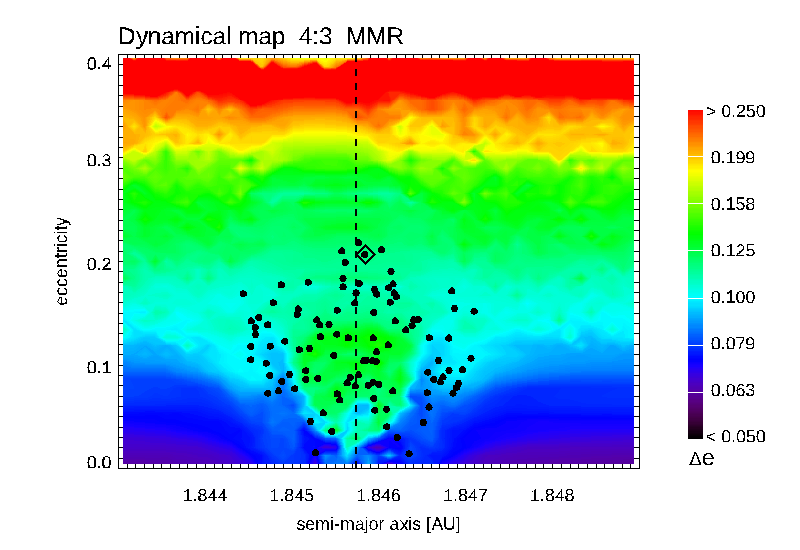}
  \includegraphics[width=0.45\linewidth]{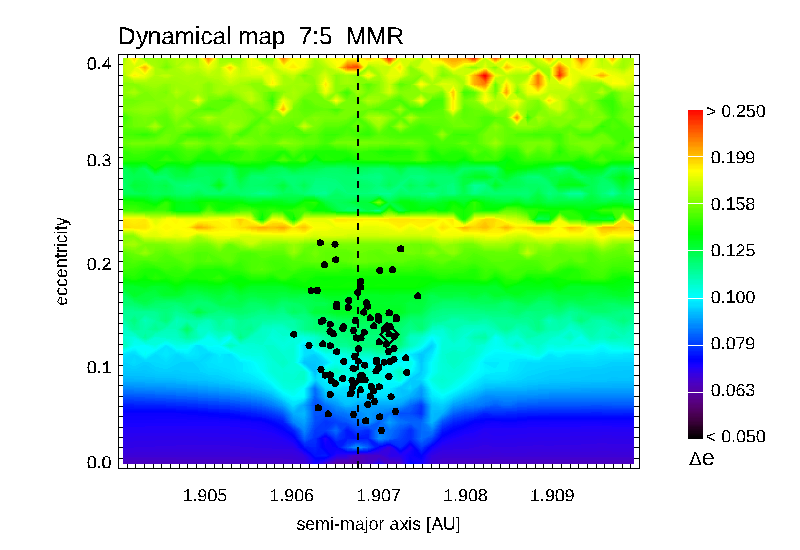}
  \includegraphics[width=0.45\linewidth]{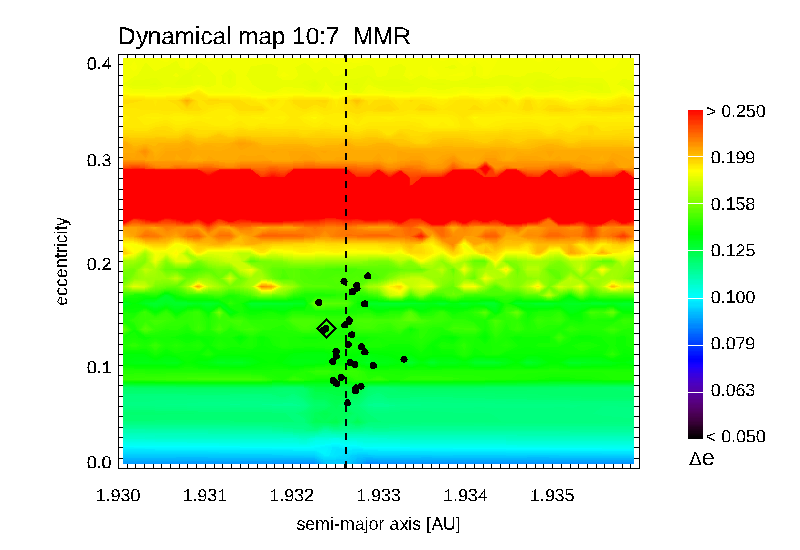}
  \includegraphics[width=0.45\linewidth]{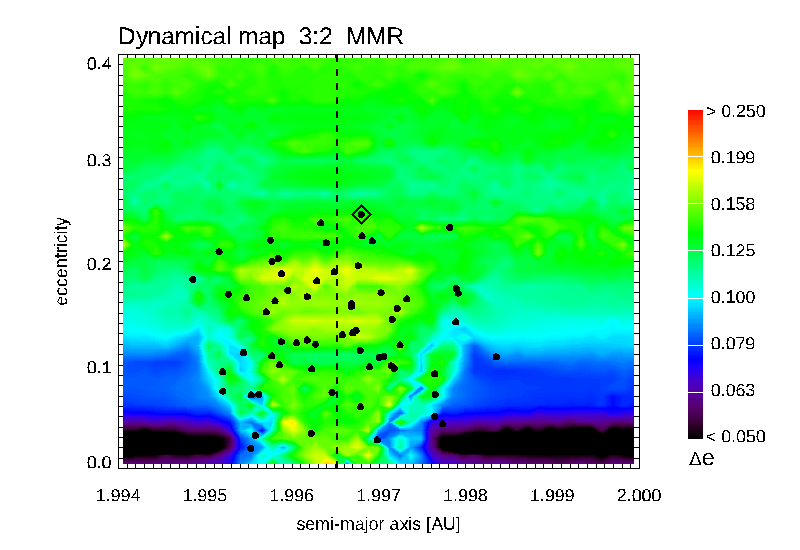}
  \caption{Dynamical maps of the four most populated MMRs using the maximum eccentricity variation as indicator. In the figures the dots show all asteroids with librating critical argument (involved in short and long term libration) in the vicinity of the exact resonance. The diamond is for the asteroid whose clones are used to compile the dynamical map of the corresponding MMR. The vertical dashed line shows the semi-major axis of the exact resonance. In order to visualise the fine structure of the dynamical map the colour scale is logarithmic, but the eccentricity values in colour bar are transformed back to linear values. When compiling the dynamical maps for the four MMRs investigated we use the same color scale that guarantees the comparability of these figures.}
\label{fig:stability-map-ae-deltaecc}
\end{figure*} 

\subsection{Dynamical maps}
In the cases presented in the previous section one can see that the asteroids belonging to the Hungaria group exhibit very interesting dynamical behaviour. In order to study the dynamical behaviour of asteroids, we compile dynamical maps of the corresponding main MMRs. First we apply the widely used method of the maximum variation of the asteroid's eccentricity. In the semi-major axis vs. eccentricity plane we assign to each $a,e$ pair the maximum of the eccentricity variation of the asteroid's orbit during the whole time length of numerical integration. In Fig.~\ref{fig:stability-map-ae-deltaecc} one can see the dynamical map of the four most populated MMRs. This figure is compiled by numerically integrating the clones of four chosen asteroids, each of them representing one MMR, in which the corresponding asteroid shows long-term libration. The initial conditions of the clones are calculated by varying the semi-major axis and eccentricity values of the given asteroid on a uniform grid (with 200 gridpoints in horizontal and 100 gridpoints in vertical direction) of the computational domain in the $a-e$ plane. The remaining orbital elements $(i,\omega,\Omega,M)$ are kept fixed to the nominal values of the original asteroid being in the MMR under study. The maximum eccentricity variation is calculated as $\Delta e_\mathrm{max} = e_\mathrm{max} - e_\mathrm{min}$, where $e_\mathrm{min}$ and $e_\mathrm{max}$ are the minimum and maximum values of the eccentricities during the one million year numerical integration. The representative asteroids of the four most populated MMRs are the following: (i) in the 4:3 MMR 2004$_{\mathrm{RG}84}$; (ii) in the 7:5 MMR 2005$_{\mathrm{QS}200}$; (iii) in the 10:7 MMR  1981$_{\mathrm{EE}17}$; and finally (iv) in the 3:2 MMR 2016$_{\mathrm{AX}18}$. In Fig.~\ref{fig:stability-map-ae-deltaecc} one can observe the fine structure of the 4:3, 7:5, and 3:2 MMRs, while the eccentricity variation does not indicate a complex dynamical behaviour in the 10:7 MMR.  

\begin{figure*}
  \centering
  \includegraphics[width=0.45\linewidth]{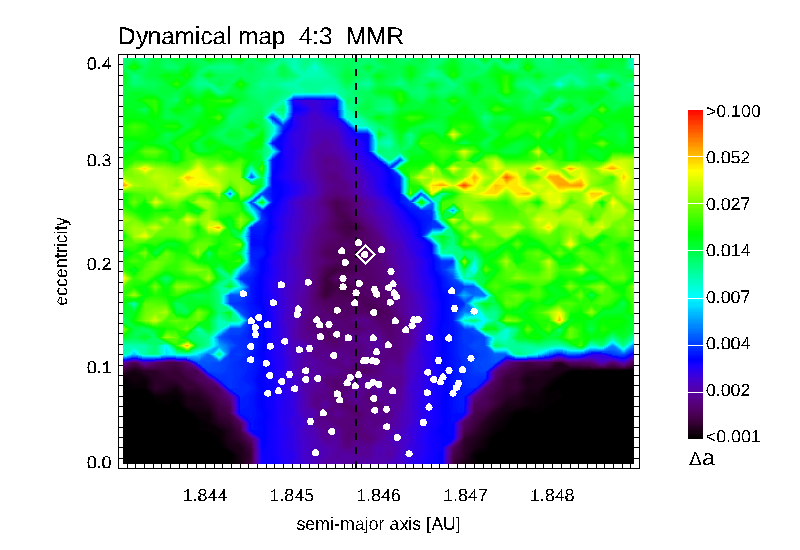}
  \includegraphics[width=0.45\linewidth]{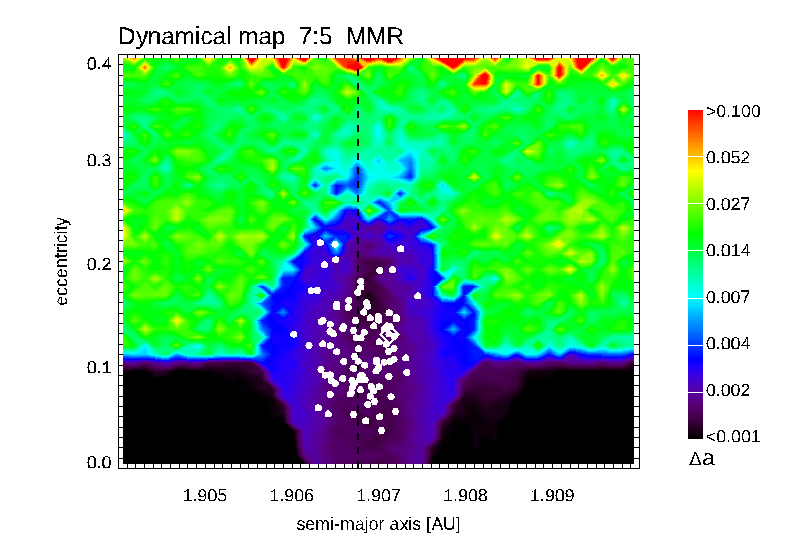}
  \includegraphics[width=0.45\linewidth]{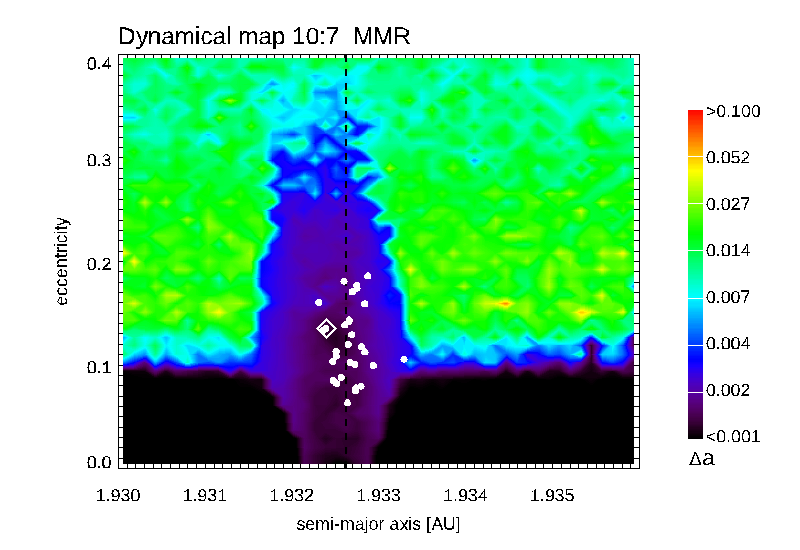}
  \includegraphics[width=0.45\linewidth]{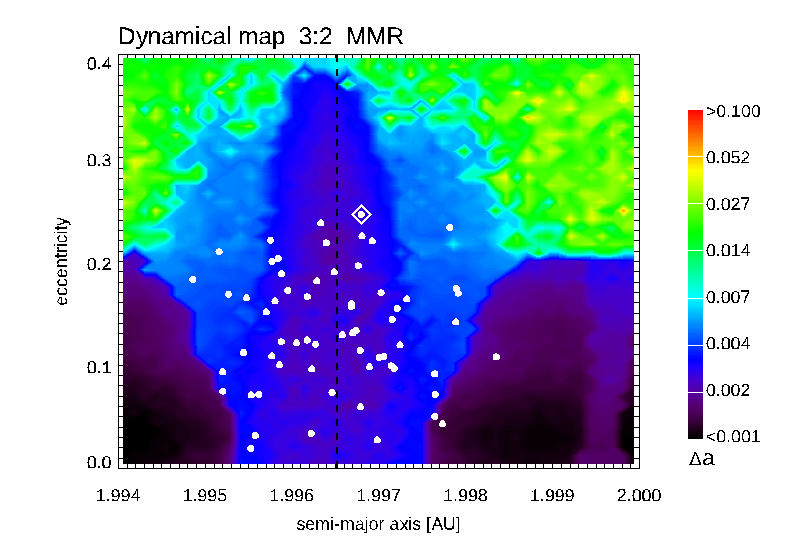}
  \caption{The same as in Fig.~\ref{fig:stability-map-ae-deltaecc}, except that the dynamical maps are calculated by using $\Delta a_\mathrm{max}$ as indicator.}
\label{fig:stability-map-ae-deltasma}
\end{figure*} 

In each case of the four most populated MMR we also display the semi-major axis and eccentricity values of asteroids yielding either short-term or long-term libration. In each figure the vertical dashed curve shows the location of the exact resonance, while the position of the representative asteroid is marked by a diamond. In order to see the resonant structure of the given MMR shown by the variation of eccentricity we use a logarithmic colour scale. The black and dark blue regions correspond to small changes in eccentricity, while the green are for the moderate, and the red regions for large eccentricity variations. Based on these figures one can see that in the neighborhood of resonances the variation of eccentricity increases. Moreover, all asteroids yielding even episodically librating critical arguments are falling to this widened (V-shaped) region although their inclination $i$, and angular orbital elements $(\omega,\Omega,M)$ differ from the corresponding orbital elements of the representative asteroid. Although the semi-major axis and eccentricity values of these asteroids are marked in these dynamical maps only for illustrative purposes, their positions fit well to the region affected by the corresponding MMR. On the other hand, to reveal the dynamics of a selected asteroid, a separate dynamical map of its neighbourhood should be calculated following the above described way.

As we already mentioned, the dynamical structure of the 10:7 MMR is not so visible than in the cases of the other three MMRs. The reason of this less detailed dynamical map could be that we use the same color scale for all of the four MMRs. The same color scale is useful for better comparison between the dynamical maps, on the other hand it could also be possible that an individual color scale for the 10:7 MMR would reveal better the fine structure of the resonance, such as the V-shape pattern.

\begin{figure*}
  \centering
  \includegraphics[width=0.45\linewidth]{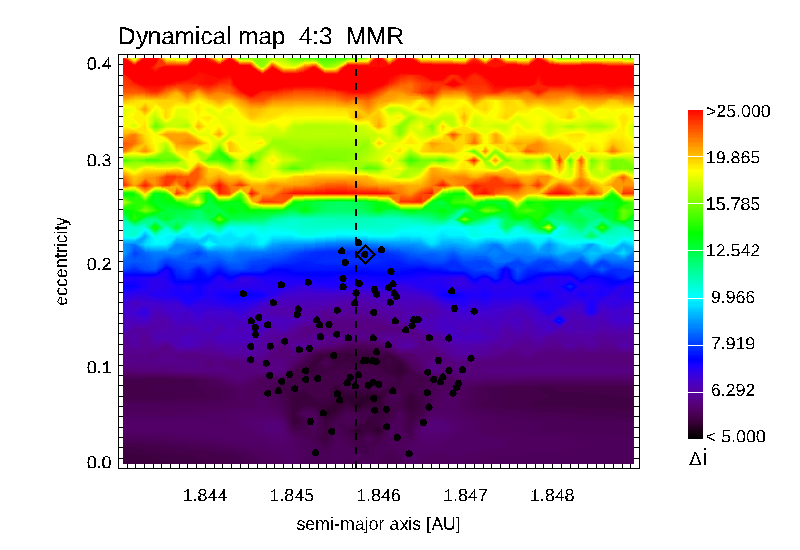}
  \includegraphics[width=0.45\linewidth]{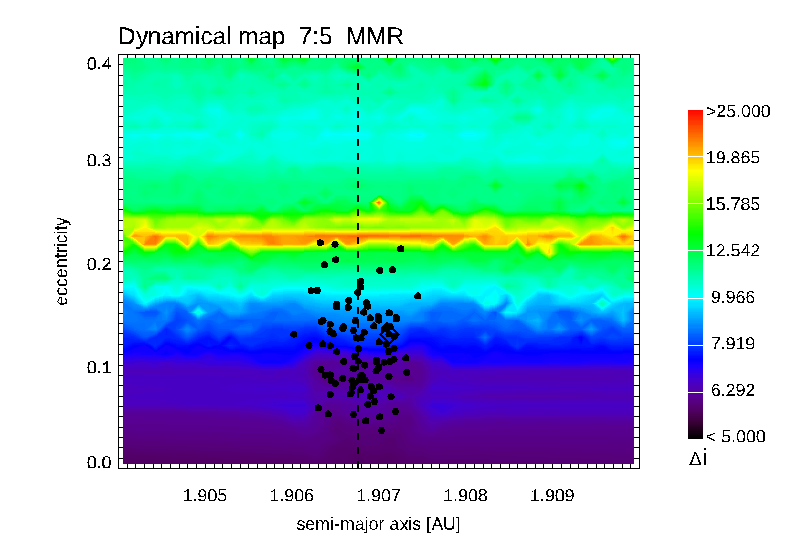}
  \includegraphics[width=0.45\linewidth]{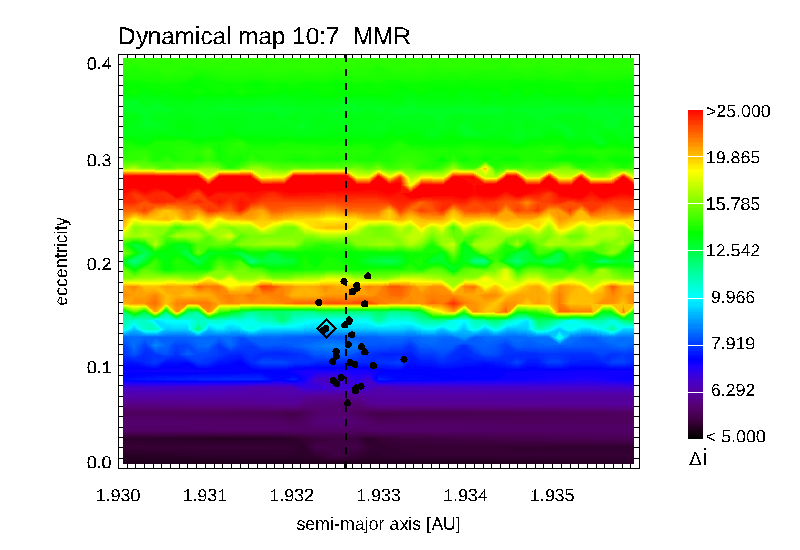}
  \includegraphics[width=0.45\linewidth]{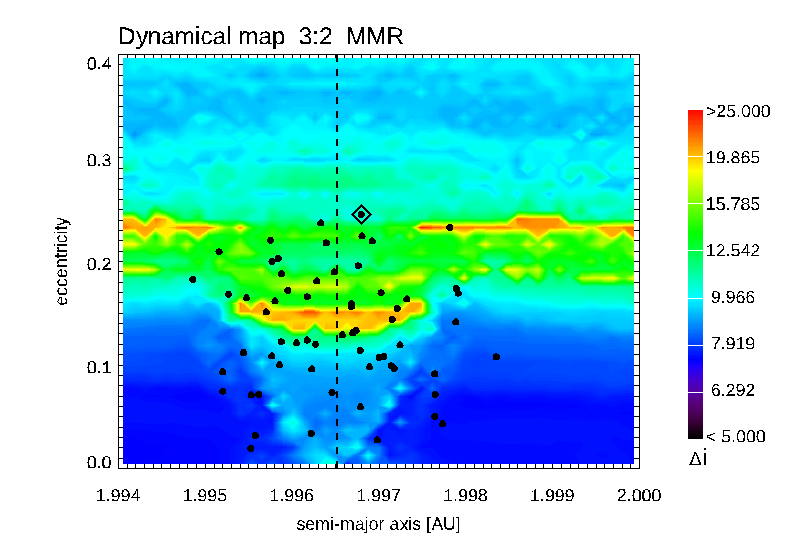}
  \caption{The same as in Fig.~\ref{fig:stability-map-ae-deltaecc}, except that the dynamical maps are calculated by using $\Delta i_\mathrm{max}$ as indicator.}
\label{fig:stability-map-ae-deltainc}
\end{figure*} 

To obtain a more complete structure about the neighborhood of a given MMR, we investigate not only the variation of the maximum eccentricity but also variations in the semi-major axis of asteroids. Using the variations in the semi-major axis as indicator is also motivated by the fact that in Figs.~\ref{fig:res-2016AX8-mmr32}-\ref{fig:res-2016UY4-mmr75} the semi-major axis shows larger oscillations around the exact position of the MMR when the critical argument circulates than when it librates. In this way, not only the variations of eccentricities, but the variations of semi-major axes can be used to compile the dynamical map of a MMR. The variation of the semi-major axis is calculated similarly to the variation of the eccentricity: $\Delta a_\mathrm{max} = a_\mathrm{max} - a_\mathrm{min}$, where $a_\mathrm{max}$ is the maximum value and $a_\mathrm{min}$ is the minimum value of the semi-major axis during the numerical integration. In Figs.~\ref{fig:stability-map-ae-deltasma} we also display the $a$ and $e$ values of asteroids with librating critical arguments and the position of the exact MMR. On the contrary to the cases when $\Delta e_\mathrm{max}$ is the tool of mapping the dynamical structure of resonances, in these figures the variation of the semi-major axis gets smaller when approaching the location of the exact resonance. Another interesting feature that can be observed in these figures is the sharp horizontal boundary at certain values of eccentricities that separates regions characterised by small and quite large semi-major axis variations. The boundary that separates small and large semi-major variations is quite sharp in the cases of the 4:3, 7:5, and 10:7 MMR being located at $e \sim 0.1$, while it is less sharp in the case of 3:2 MMR, and it is located at $e \sim 0.2$.

\begin{figure}[!t]
   \includegraphics[width=0.9\linewidth]{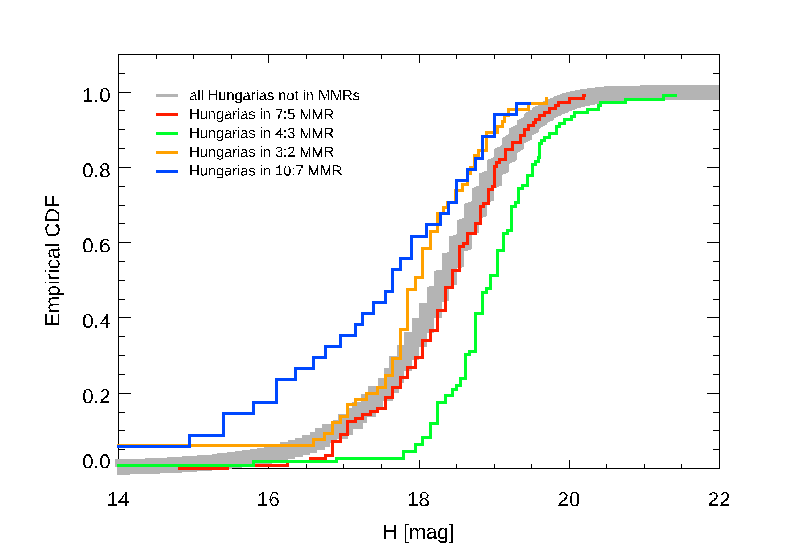}
   \caption{Empirical cumulative distribution function for absolute magnitude of the asteroids captured in the four main MMRs and not captured in MMRs.}
   \label{fig:CDF}
\end{figure} 

\begin{figure}[!t]
   \includegraphics[width=0.9\linewidth]{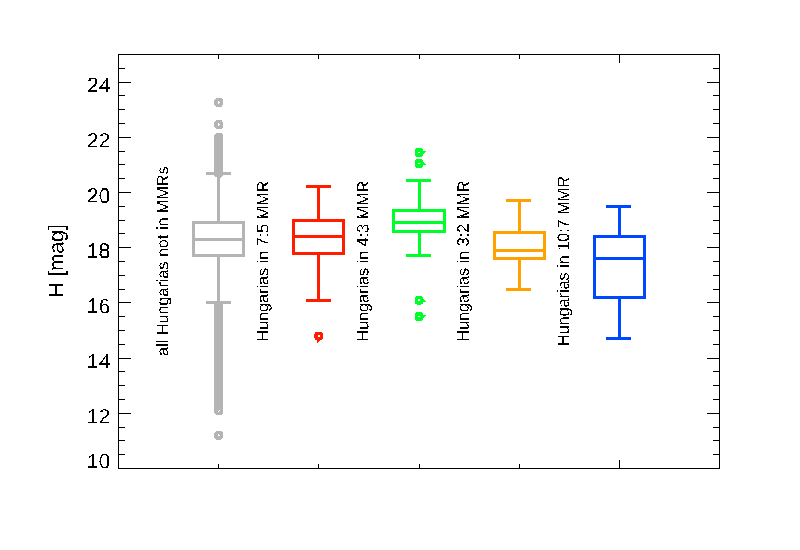}
   \caption{Box and whiskers plot from absolute magnitude of the asteroids captured in the four main MMRs and not captured in MMRs. The box encloses the interquartile range (IQR), defined at IQR75-IQR25. The whiskers extend out to the maximum or minimum value of the data, or to the 1.5 times either the IQR75 or IQR25, if there is data beyond this range. Outliers are identified with small circles.}
   \label{fig:Boxplot}
\end{figure} 

The dynamical maps of a certain MMR show different structures when using different indicators ($\Delta a_\mathrm{max}$ versus $\Delta e_\mathrm{max}$). For instance, in the previous dynamical map we find that around the MMR the values of $\Delta e_\mathrm{max}$ are increased, while the values of $\Delta a_\mathrm{max}$ are decreased. For lower eccentricity values of the asteroids the V-shape pattern is also recognisable, while for larger eccentricities the region affected by the resonance shrinks. The overall resonant pattern takes the form of a deltoid. The shrinking of the domain affected by the resonance cannot be observed in the dynamical maps that are compiled by using $\Delta e_\mathrm{max}$ as indicator. An additional feature of the dynamical map around the 4:3 MMR is that the upper part of the deltoid is bending towards smaller values of semi-major axis. Already in dynamical maps obtained by $\Delta e_\mathrm{max}$ can be observed that the 3:2 MMR is extended in a wider region than the other three resonances, this difference is even more stringent when using $\Delta a_\mathrm{max}$. Moreover, by using $\Delta a_\mathrm{max}$ as indicator one can conclude that the 3:2 MMR is not as stable as the three other MMRs regarding the variations in semi-major axis. We note that the bending deltoid structure we observe in this case is already found in a study \citep[see][]{2012CeMDA.113...95E} when the dynamical maps of some higher order MMRs of outer type have been computed by using the RLI (Relative Lyapunov Indicator) chaos indicator \citep{2004CeMDA..90..127S} and the librating critical argument.

Motivated by the fact that by introducing $\Delta a_\mathrm{max}$ as indicator, we also apply the maximum inclination variation $\Delta i_\mathrm{max} = i_\mathrm{max} - i_\mathrm{min}$. The results obtained are displayed in Fig.~\ref{fig:stability-map-ae-deltainc}, where we find further interesting dynamical features around the MMRs. One of the most stringent features is the sudden change in inclination variation at certain values of the eccentricities of asteroids. A deeper study, which is planned in a forthcoming paper, is needed to decide whether these structures are due to secular resonances or close encounter with Mars. Regarding their structures the four main MMRs show similarities, however in the case of the 3:2 MMR the maximum variation of the inclination is larger than in the cases of the three other cases. When comparing the dynamical maps obtained by $\Delta a_\mathrm{max}$ and $\Delta i_\mathrm{max}$, one can conclude that the 3:2 MMR might be less stable than the three other main MMRs.

\subsection{The overall dynamical state of Hungaria asteroids using real data as initial conditions}

\begin{figure*}
  \centering
  \includegraphics[width=0.45\linewidth]{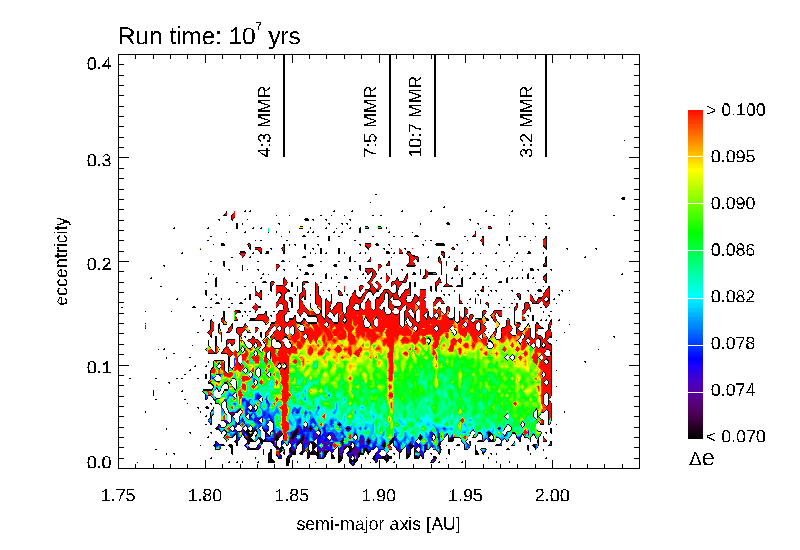}
  \includegraphics[width=0.45\linewidth]{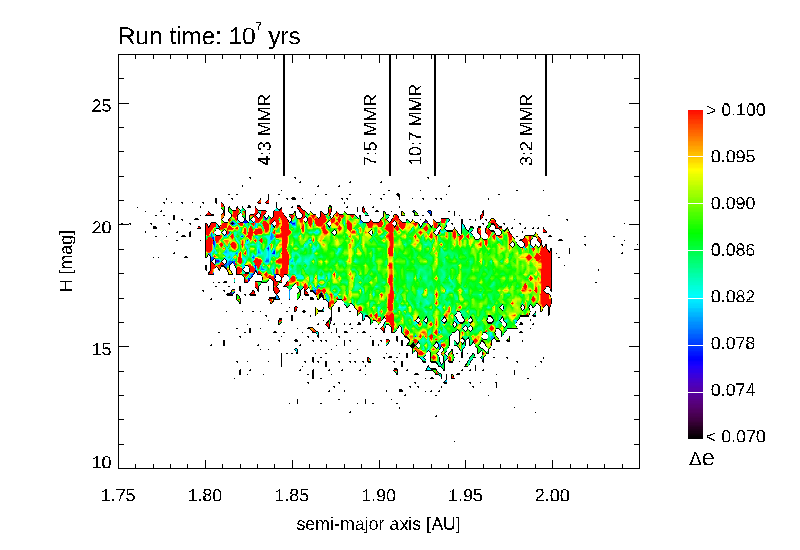}
  \includegraphics[width=0.45\linewidth]{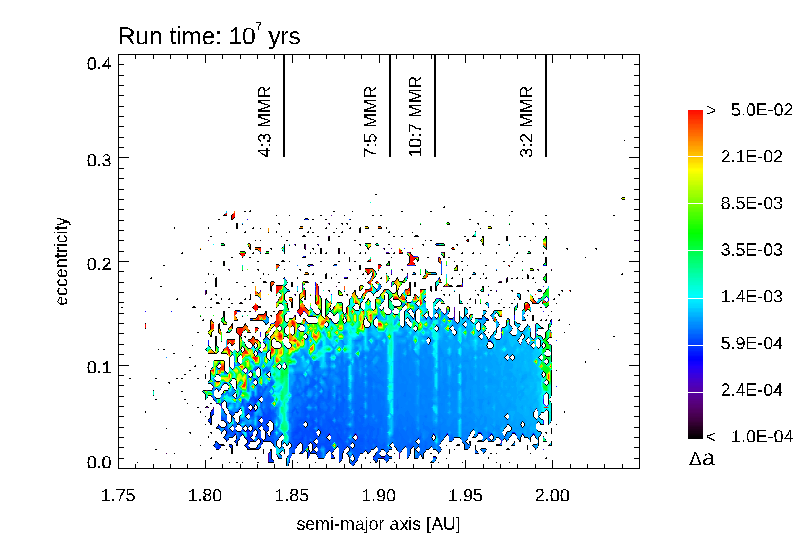}
  \includegraphics[width=0.45\linewidth]{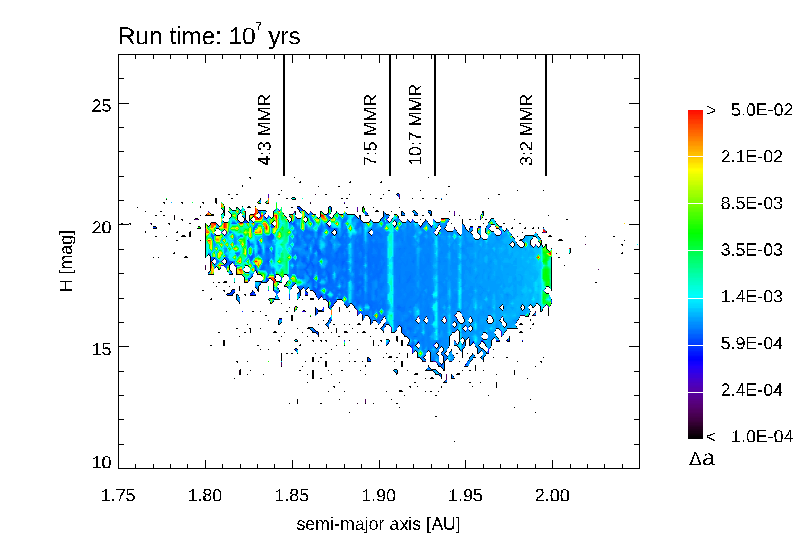}
  \includegraphics[width=0.45\linewidth]{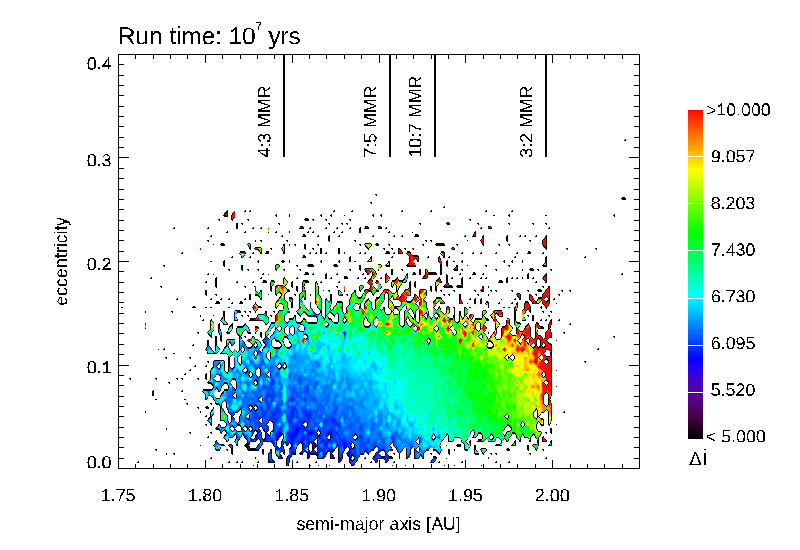}
  \includegraphics[width=0.45\linewidth]{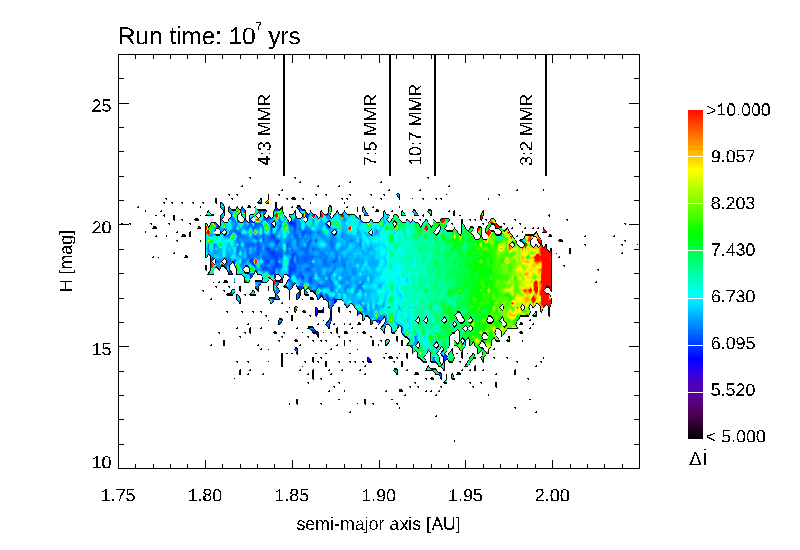}
  \caption{Dynamical state of the whole Hungaria group at a given epoch. We use the maximum variations of eccentricity, semi-major axis, and inclination as dynamical indicators. We show the dynamical state of the whole Hungaria group on the $a-e$ and $a-H$ parameter planes in the right and left panels, respectively. Instead of test particles taken from a regular grid, the orbital elements and absolute magnitudes of real asteroids are used.}
\label{fig:map-hun}
\end{figure*}

In our investigations so far we map the structure of the four main MMRs by cloning the representative asteroids of the given MMR obtaining large swarms of test particles. Our next question is, how the dynamical map would be modified if instead of test particles we numerically integrate the orbits of real asteroids belonging to the group of Hungaria asteroids, and use the previously introduced indicators of dynamical behaviour. A mapping of the dynamical state of the Hungaria asteroids at a given epoch can be done because the number of the known Hungaria asteroids increased a lot due to the recent sky surveys, enabling the use of their orbital data as initial conditions of numerical integrations. To do so, we divided the computational domain of the semi-major axis ($a=1.75 - 2.05$ au) to 200 cells, the computation domain of the eccentricity ($e=0-0.4$) to 100 cells and the computational domain of the absolute magnitude ($H=10-27$) to 100 cells. In the following, we count all asteroids whose semi-major axis and eccentricity or absolute magnitude values fall in the given cell, and the median value of their $\Delta a_\mathrm{max}$, $\Delta e_\mathrm{max}$, and $\Delta i_\mathrm{max}$ are determined, respectively. The cell is then coloured according to these median values in the corresponding panels of Fig.~\ref{fig:map-hun}. Our approach is definitely a novel one when comparing it to the usual dynamical maps where the action-like initial orbital elements are taken from a regular grid either from the $(a,e)$ and the $(a,i)$ planes, while the angular elements are kept fixed to certain values. We should also note that our novel approach is only applicable when in the investigated region of the parameter space asteroids are present in large number, and are not emptied due to strong destabilization mechanisms such as strongly chaotic behaviour due to close encounters.

Our results are shown in the six panels of Fig.~\ref{fig:map-hun}, in which $\Delta e_\mathrm{max}$, $\Delta a_\mathrm{max}$, and $\Delta i_\mathrm{max}$ are displayed both in the $(a,e)$ and $(a,H)$ parameter planes. The dynamical structure shown in these panels are obtained after 10 million years long numerical integration. In those panels that have been compiled by using $\Delta e_\mathrm{max}$ and $\Delta a_\mathrm{max}$ the MMRs are visible very well. It is noteworthy that on the upper left panel of Fig.~\ref{fig:map-hun}, in which the $(a,e)$ plane is mapped, the eccentricities of the asteroids begin to show larger variation already when $e \sim 0.12$, which is clearly less than the maximum eccentricity limit of Hungaria asteroids. The mechanism that is responsible for the larger eccentricity variations of asteroids is the overlapping of the MMRs with Mars, a phenomenon found and described among Hungaria asteroids recently by \cite{2018MNRAS.479.1694C}.

In the middle panels we use $\Delta a_\mathrm{max}$ as indicator, and one can see that even the weaker MMRs can be identified, because the overall variation of the semi-major axes around the MMRs is larger than the surrounding regions of the $a-e$ plane. As a general conclusion, by investigating carefully Fig.~\ref{fig:map-hun} one can conclude that the MMRs of outer type with Mars result in larger variations in the eccentricities and semi-major axes of the majority of real asteroids involved in those resonances.

When $\Delta i_\mathrm{max}$ is applied as indicator an interesting phenomenon can be observed: at smaller semi-major axes the variation of the inclinations of the real asteroids are also smaller than at larger values of semi-major axes. Although a few of the MMRs can be identified also in this case, there is a general trend in increasing of the inclinations towards larger semi-major axes, therefore the variations of the inclinations are suppressed in the figure. Thus $\Delta i_\mathrm{max}$ among the real Hungaria asteroids is not so strongly influenced by the different MMRs with Mars see the lower panels of Fig.~\ref{fig:map-hun}.

Summarizing our findings from Fig.~\ref{fig:map-hun} we can conclude that the most visible dynamical features on these panels are due to the various MMRs with Mars. This does not mean that other effects do not shape the dynamics of Hungaria asteroids, but the most visible ones are certainly due to the MMRs.




The presented dynamical investigations reveal the very complex and interesting dynamics of asteroids in the Hungaria group due to the MMRs with Mars. Beside, we find that for calculation of dynamical maps around various MMRs, the application of the maximum variation of semi-major axes is also a suitable tool, that in certain cases could reveal the fine structure of the MMRs investigated. The applicability of $\Delta a_\mathrm{max}$ as dynamical indicator is based on the fact that the larger the variation of the semi-major axis, the larger the libration amplitude of the critical argument.

\section{Comparison of the absolute magnitudes and dynamical properties of asteroids}

As a last step of our survey, we investigate possible relationships between dynamical and physical properties of the asteroids belonging to the Hungaria group. There has been a series of measurements \citep{2017Icar..291..268L,2019Icar..322..227L} aiming at revealing various physical properties of Hungaria asteroids such as the absolute magnitude, size, shape, and albedo. These measurements are thought to help discover the variety of asteroids, in a broader sense even the properties of the building blocks of the ancient Solar System. 
In our work we are going to study the absolute magnitude of the Hungaria asteroids in the context of their dynamics. Our choice is motivated by the fact that the absolute magnitude can be determined for all asteroids, while their other physical parameters are hardly known. On the other hand, the absolute magnitude of an asteroid also depends on many other physical properties. We may know the spectral properties of an asteroids and therefore its surface composition, however, the surface forming processes can affect the asteroid's albedo and therefore its absolute magnitude, see for details in \cite{2010Icar..209..564G}. For instance, one can consider different kind of space weathering effects that result in different albedo and spectral properties for objects originally with similar surface composition. Thus space weathering is a significant source of uncertainty when performing a taxonomic classification.  

In Fig.~\ref{fig:CDF} we show the empirical cumulative distribution function (CDF) of the absolute magnitudes  calculated for Hungaria asteroids. The thick grey line is calculated for those asteroids that are not affected by any MMR. The coloured lines are calculated for asteroids influenced by the four most populated MMRs. There are two significant differences with respect to the grey line: (i) asteroids found at the 10:7 MMR are brighter than the average value of Hungarias, (ii) while asteroids around the 4:3 MMR are fainter. Beside these two very stringent features, in the case of asteroids at the 3:2 MMR one can also see a small deviation from the average magnitude values in the case of fainter bodies.

To better visualize the above described properties in the distribution of the absolute magnitudes, we also show box and whiskers plot in Fig.~\ref{fig:Boxplot}. We recall the reader that a box plot is a way to graphically depict different groups of numerical data using their quartiles. Whiskers are lines that extend outside of boxes indicating the variability of data outside the lower and upper quartiles. Outlying data are plotted as separate points. Box plots are used to display variation in the given samples without any a priori assumption or knowledge on their statistical distribution. The spacings between the different parts of the box show the degree of dispersion and the possible skewness in the data, too.

Studying Fig.~\ref{fig:Boxplot} one can also see that asteroids in the 4:3 MMR are fainter than asteroids in the 10:7 MMR. Moreover, the distribution of asteroids in the 4:3 MMR is symmetric, while the distribution of asteroids in 10:7 MMR is elongated towards the brighter bodies, in other words the distribution has a negative skew. In the case of the asteroids in the 3:2 MMR, the distribution of asteroids had a positive skew that is the reason of the deformity in the empirical CDF figure, see Fig.~\ref{fig:CDF}, meaning that in the sample a considerable number of fainter objects can be found.

The above described results can be explained, however, with strong assumptions. In the following, we describe a possible scenario but with emphasizing its uncertainty due to the lack of direct measurements concerning the surface of these bodies. 

By assuming a similar albedo for these asteroids, based on Figs.~\ref{fig:CDF} and Fig.~\ref{fig:Boxplot} asteroids in the 10:7 MMR might be larger, while asteroids in the 4:3 MMR might be smaller in size than the other ones. This assumption is also supported by Fig.~\ref{fig:sma-H}, where the sign of the Yarkovsky effect is clearly seen. Since the Yarkovsky effect is weaker for larger bodies, moreover a MMR can block the radial drift of the affected bodies \citep[see][]{2015aste.book..509V}, one can conclude that in the 10:7 MMR are really larger bodies involved. On the contrary, the Yarkovsky effect is stronger for smaller asteroids resulting in a longer range radial drift from the original location of the catastrophic collision. This might be the reason that in the 4:3 MMR smaller asteroids can be found. 

We would like to emphasize that the aim of our investigation is not the determination of the physical characteristics of Hungaria asteroids, we are seeking for eventual connections between their physical and dynamical properties. Our results show that a given MMR becomes significant when a large number of bodies are drifting to the nominal locations of the MMRs. The reason of such an orbital drift might be a scattering effect due to the collision event itself, and later on the Yarkovsky effect.  

We are aware the fact that the above described scenario might be valid only for the members of the collisional family, as they show the V-shape pattern on the $a-H$ plane. On the other hand, it is not uncommon in the literature that the whole Hungaria group is considered in dynamical studies, \citep[e.g.][]{2009Icar..204..172W}, since the identification of the family members can be cumbersome and uncertain. However, our speculative scenario might not be restricted to the members of the collisional family, because the Yarkovsky effect acts on the whole group of asteroids. Depending on their spin, some of the group members are drifting inwards, some of them are drifting outwards, which in turn would result the observed distribution of asteroids. During their drift, as has been found by \citet{2010Icar..210..644M}, asteroids can be captured into various MMRs, and escape from the swarm of Hungaria asteroids due to the increase of their eccentricities up to the Mars-crossing limit. Moreover, according to \citet{2010Icar..210..644M}, a repopulation of the Hungaria asteroid region is also possible through this escape mechanism, as Hungaria asteroids escaping from the MMRs can undergo close encounters with Mars, and could be injected back to the region of Hungaria asteroids. It is clear therefore, that Hungaria asteroids are not moving only under the Yarkovsky effect, they may undergo drastic orbital changes due to the very complex gravitational perturbations acting on them. On the other hand, the observed distribution of asteroids requires a certain size segregation process that might be most easily explained by the Yarkovsky effect. A more careful analysis aiming at revealing the size differentiation process of asteroids is out of the scope of this work, and will be the subject of a forthcoming study.

\section{Conclusion}\label{sec:conclusion}

Asteroids in the Hungaria group are of high interest in both dynamical studies since chaotic diffusion and non-conservative effects play important role in shaping their long-term dynamics, and in investigating their physical properties to obtain more knowledge on the ancient planetesimal population that might have contributed to the formation of terrestrial planets. These two very important research areas can be observed in several studies in which the dynamical stability of the region occupied by Hungaria asteroids has been addressed \citep[e.g.][]{2010Icar..207..769M,2018MNRAS.479.1694C}, moreover, due to the recent observations, there is a growing knowledge about the physical properties of Hungaria asteroids, too \citep[e.g.][]{2017Icar..291..268L,2019Icar..322..227L}. 

In this paper we have performed a comprehensive survey on the Hungaria group of asteroids. By using the method FAIR \citep{2018MNRAS.477.3383F}, considering all known asteroids belonging to the Hungaria group, we have identified several MMRs between these bodies and the planet Mars. According to our results, the four most populated MMRs in this region are the 4:3, 7:5, 10:7, and 3:2 ones. In the cases of these MMRs we have performed detailed dynamical investigations, in which beside the classical maximum eccentricity variation method, we have calculated the maximum variation of other action-like orbital elements such as the asteroids' semi-major axes and inclinations. Our results reveal the complex dynamical properties of the regions around the above mentioned MMRs also indicating the usefulness of simultaneous and complementary application of the maximum variations of $a, e, \mathrm{and}\ i$.

According to our dynamical investigations, there are two types of resonant behaviour of asteroids, namely, when the critical argument (resonance variable) shows short-term or long-term libration as has been found also among Hungaria asteroids by \cite{2021Icar..36714564C}. It can also happen that between two short-term librations there is a period of circulation. The short-term libration of the critical argument is more frequent than the long-term one being a clear sign of chaotic behaviour. We observed in the case of almost all MMRs that during the libration of the critical argument the eccentricity of an asteroid is changing between higher limits, while in the case of circulation the eccentricity becomes lower (cf.~Fig.~\ref{fig:res-2016AX8-mmr32}--\ref{fig:res-2016UY4-mmr75} in right panels). Dynamical maps of the four most populated MMRs, and the map of the dynamical state calculated by using the orbital data of real Hungaria asteroids for the initial conditions, reveal the high importance of the MMRs with Mars. These MMRs dynamically excite the involved asteroids, their overlapping determines the eccentricity limit for the asteroids, moreover MMRs can act as leaking chanels for the members of the Hungaria population that are affected.

According to the above mentioned results, the members of the Hungaria groups are ideal targets of studying chaotic diffusion that is a very important feature of the dynamics of asteroids \citep{2007LNP...729..111T}, and the effects of non-conservative dynamics such as the Yarkovsky/YORP effects. Another very interesting question is the existence of the outer sharp boundary of the Hungaria group at the 3:2 MMR, while the inner boundary of the population cannot be related to the 4:3 MMR, because asteroids seem to be drifting inward through this MMR and their number is gradually decreasing towards smaller semi-major axes. It would be noteworthy a thorough investigation of the long-term dynamics of asteroids at the outer boundary of Hungaria population to reveal the possible role of the 3:2 MMR in destabilization of Hungaria asteroids.

During our statistical investigations, we have found an interesting relationship between Hungaria asteroids that might be certainly the results of the Yarkovsky effect. Near to the semi-major axis of (434) Hungaria, there is the location of the 10:7 MMR. Asteroids involved in this MMR are more brighter than the average brightness of the group members, while asteroids affected by the 4:3 MMR are fainter than the average value. This latter property is very interesting, since the nominal position of the 4:3 MMR is farther from semi-major axis of (434) Hungaria asteroid that is supposed to be the parent body of the Hungaria collisional family. Assuming that the members of the collisional family have similar albedo (that is an acceptable assumption, because the surface albedo reflects the minearological composition of the progenitor body), asteroids involved in the 10:7 MMR should be bigger in size, and therefore the Yarkovsky effect on them is weaker, while the smaller family members involved in the 4:3 MMR could have been drifted quite far away from the position of the catastrophic collision due to the stronger Yarkovsky effect acting on them. Due to this latter effect, smaller asteroids are drifted through longer radial range, and are being involved in the 4:3 MMR later. This finding is supported by Fig.~\ref{fig:sma-H}, too. On the other hand, one should be careful when making the above assumptions, because Hungaria asteroids are also the subject of a complex dynamical behaviour due to the MMRs with Mars, and secular resonances. Moreover, space weathering can alter the surface properties of asteroids. The relationship between albedo, spectral properties, and absolute magnitude of asteroids is hardly discovered yet. To understand better the various space weathering processes in situ measurements of the surface of asteroids or even sample return missions would be of high importance \citep{2010Icar..209..564G}.

As we already mentioned, the Hungaria family is formed presumably as the result of a catastrophic collision, in which one of the involved bodies is (434) Hungaria asteroid. The resulting family members therefore inherit the physical and dynamical properties of the parent bodies, therefore it is of high importance to know the properties of the parent bodies. Such a collision event might be modelled by Smooth Particle Hydrodynamic simulations \citep[e.g.][]{2013AN....334..996M, 2019Icar..317..215J} combined with a simple numerical treatment of the Yarkovsky effect. Although these simulations are very time consuming, and there could be many uncertainties even in the physical models applied, they could be useful for better understanding the formation and origin of the Hungaria collisional family. As a future work, we intend to study the formation and evolution of Hungaria asteroids numerically, including their collisional formation and also the effects of non-conservative forces, moreover, the chaotic diffusion shaping their dynamical properties.

\begin{acknowledgements}
We thank the reviewer Dr. J. Correa-Otto for the useful comments and suggestions that helped us improve the manuscript. EF-D and ZsS thank the support of the Hungarian National Research, Development and Innovation Office (NKFIH), under the grant K-119993.  We acknowledge the computational resources for the Wigner GPU Laboratory of the Wigner Research Centre for Physics.
\end{acknowledgements}

%
   \bibliographystyle{aa} 
   \bibliography{main} 
%

\end{document}